\documentclass[10pt,a4paper]{article}

\usepackage{amsmath,amssymb,amsthm}
\usepackage{url}
\usepackage{placeins}
\usepackage{subfigure}
\usepackage{multirow}
\usepackage{epsfig}
\usepackage{natbib}
\usepackage{rotating}

\newcommand{\Vhat}{\ensuremath{{\widehat{V}}}}
\newcommand{\betaBar}{\ensuremath{\bar{\beta}}}
\newcommand{\DeltaBar}{\ensuremath{\bar{\Delta}}}

\newcommand{\I}[1]{\ensuremath{ I_{#1}  }}

\newcommand\ben{\begin{equation}}
\newcommand\een{\end{equation}}
\newcommand\bea{\begin{eqnarray*}}
\newcommand\eea{\end{eqnarray*}}
\newcommand\bean{\begin{eqnarray}}
\newcommand\eean{\end{eqnarray}}

\begin{document}
\author{Chris Kenyon\footnote{Contact: chris.kenyon@lloydsbanking.com}  and Andrew Green\footnote{Contact: andrew.green2@lloydsbanking.com}}
\title{Warehousing Credit (CVA) Risk,\\ Capital (KVA) and Tax (TVA) Consequences\footnote{\bf The views expressed are those of the authors only, no other representation should be attributed.  The authors are not tax experts and this does not constitute tax advice of any kind.  Not guaranteed fit for any purpose.  Use at your own risk.}}
\date{06 January 2015\\ \vskip5mm Version 2.00}

\maketitle

\begin{abstract}
Credit risk may be warehoused by choice, or because of limited hedging possibilities.  Credit risk warehousing increases capital requirements and leaves open risk.  Open risk must be priced in the physical measure, rather than the risk neutral measure, and implies profits and losses.  Furthermore the rate of return on capital that shareholders require must be paid from profits.  Profits are taxable and losses provide tax credits.  Here we extend the semi-replication approach of Burgard and Kjaer (2013) and the capital formalism (KVA) of Green, Kenyon, and Dennis (2014) to cover credit risk warehousing and tax, formalized as double-semi-replication and TVA (Tax Valuation Adjustment) to enable quantification.
\end{abstract}

\section{Introduction}

Credit valuation adjustments (CVA) apply to all counterparties with derivatives transactions that are marked to market, that is, those in the Trading Book.  For most banks only a subset of these counterparties have liquid CDS contracts available for hedging default risk, so some credit risk is inevitably warehoused.  Increased credit risk requires increased capital.  Open risk requires pricing in the physical measure rather than the risk neutral measure.

Here we extend the semi-replication approach in \citet*{Burgard2013a} and \citet*{Green2014b} to include counterparty credit risk warehousing and taxation of any resultant profit or loss.  Thus we introduce double-semi-replication, that is, partial hedging of value jump on counterparty default, and Tax Valuation Adjustment(TVA).  

Credit risk also affects the capital a bank is required to hold and we include this interaction.  In addition shareholders, and bank executives, require a return on this deployed capital, which we have previously formalized within KVA \cite{Green2014b}.  Paying a return on capital requires profits and so we  include the taxation of these profits within our double-semi-replication approach.

\subsection{Limits of CVA Hedging}
The USD denominated liquid CDS market contains only around 1600 names.
Banks typically have thousands to tens of thousands of counterparties so relatively few will be present in the liquid CDS market.  Outside the USD market there are relatively few liquid CDS contracts.  CDS  markets, even when nominally liquid, may be shallow \cite{Carver2013c} making pricing using observed spreads problematic and hedging challenging. 




Some desks may warehouse CVA risk by choice.  The pricing effect w.r.t. the bank is the same as though the warehousing was driven by lack of liquid CDS, except that there may be better data to estimate hazard rates. Risk must be priced in otherwise pricing will be inconsistent with the risk appetite of the bank.  Here we capture this risk appetite using a market price of jump risk \cite{Antje2005a,Berg2010a}, which can be calibrated to the bank price of jump risk.   Depending on the sign of the price of jump risk this may be higher or lower than the price of fully hedged CVA. 

Computing market-implied hazard rates from CDS spreads is problematic because credit protection also confers capital relief which may be priced in \cite{Kenyon2013d}.  Thus CDS spreads can only be used to obtain bounds on market-implied hazard rates.  However, by Regulation, observed CDS spreads must be used for computing CVA VAR capital further complicating the picture.

\subsection{Taxation and TVA}
Whilst default has been widely considered in the derivatives literature, tax has so far been almost absent outside Corporate Finance and the Banking Book (\cite{Kenyon2013a} is one exception).  Tax is potentially present whenever there is a profit or loss so when a derivative is perfectly hedged tax can be ignored. However, as has been demonstrated above, CVA trading desks cannot in practice hedge all of their default risk and as a consequence profits and losses must occur.  Profits are taxable and losses create so-called tax shields.  Here we formalize a \emph{Tax Valuation Adjustment} (TVA\footnote{TVA has been used before in XVA as ``Total Valuation Adjustment'', but given that both KVA and Tax effects were not included we are reusing the acronym.}) to quantify these effects when hedging is not perfect. 

TVA is the valuation adjustment that formalizes the effect of tax on profits and losses.  That is, if a return on capital is required by shareholders then this must be paid out of profits, and profits in this instance are after-tax.  Thus we must allow for tax when calculating the price required for a derivatives such that we can pay shareholders for capital.  Equally, when the bank has expected losses from counterparty jump to default because of limited hedging then the bank will require compensation in order to do the trade.  We choose this compensation to be equal in physical-measure expectation to the expected unhedged losses.  Now, whilst the counterparty has not defaulted this cashflow stream may be regarded as profit by the tax authority.  If the counterparty defaults then there will also be a tax consequence on the default date.  Default losses will provide a tax credit that will offset the previous tax costs on these profits.  

\subsection{Capital}
Shareholders, and bank executives, require a return on deployed capital.  This return is paid out of profits and hence is subject to taxation.  There is a whole literature on methods to return profits to shareholders using share buy-backs or dividends.  These methods have different tax consequences {\it to the shareholders} and hence are out of scope here.  We focus on the tax consequences {\it to the bank} of profits to pay the return on capital, as well as from the profits and losses because of partial hedging of jump to default credit risk.

\subsection{CVA Hedging and CVA Regulatory Capital}
Under Basel III \cite{BCBS-189} capital relief is granted for CVA hedging using single name and index CDS under both standard and advanced calculation methods. If a CDS is eligible as a hedge under regulatory capital rules and the CVA trading desk chooses not to fully hedge the default risk then the capital requirement will be higher than it would otherwise be. This clearly has direct implications for KVA.   Here we establish the relationship between CVA risk warehousing, KVA and TVA.  

\section{Warehousing CVA Risk and Double Semi-Replication}
\begin{table}
\centering
\small
\begin{tabular}{|p{3.5cm}|p{10cm}|}\hline
{\bf Parameter} & {\bf Description}\\\hline 
$\Vhat(t, S)$ & The economic value of the derivative or derivative portfolio\\
$V$ & The risk-free value of the derivative or derivative portfolio\\ 
$U$ & The valuation adjustment\\
$X$ & Collateral\\
$K; K_U; K_R$ & Total Capital Requirement; Unhedged Capital Requirement; Capital Relief from full CDS hedging\\
$E$ & Cash flow liable to tax as a function of time\\
$\Pi$ & Replicating portfolio\\
$S$ & Underlying stock\\
$\mu_S$ & Stock drift\\
$\sigma_S$ & Stock volatility\\
$P_C$ & Counterparty Bond (zero recovery)\\
$P_1;\ P_2$ & Issuer bond with recovery $R_1$; recovery $R_2$, note $R_1\ne R_2$\\
$d\bar{\beta}_S$;$\ d\bar{\beta}_C$;$\ d\bar{\beta}_X$;$\ d\bar{\beta}_K$;$\ d\bar{\beta}_E$ & Growth in the cash account associated with stock; counterparty bond; collateral; capital; tax. All prior to rebalancing.\\
$r;\ r_C;\ r_i;\ r_X;\ r_F$ & Risk-free rate; Yield on counterparty bond; issuer bond; collateral;  issuer bond (one-bond case)\\
$M_B;\ M_C$ & Close-out value on issuer default; Counterparty default\\
$\alpha_C;  \alpha_i$ & Holding of counterparty bonds with full hedge $(\psi=1)$; issuer bond\\
$\delta$ & The stock position\\
$\gamma_S$ & Stock dividend yield\\
$q_S;\ q_C$ & Stock repo rate; counterparty bond repo rate\\
$J_C;\ J_B$ & Default indicator for counterparty; issuer\\
$g_B;\ g_C$ & Value of the derivative portfolio after issuer default; counterparty default\\
$R_i;\ R_C$ & Recovery on issuer bond $i$; counterparty derivative portfolio\\
$\lambda_C;\ \lambda_C^{\mathbb{P}};\ \tilde{\lambda}_C;\ \lambda_B$ & Effective financing rate of counterparty bond $\lambda_C= r_C - r = r_C - q_C$; Physical measure hazard rate for counterparty C; Effective hazard rate under semi-replication; Spread of a zero-recovery zero-coupon issuer bond. For bonds with recovery the following relation holds $(1 - R_i)\lambda_B = r_i - r$ for $i\in\{1,2\}$\\
$s_F;\ s_X$ & Funding spread in one bond case $s_F = r_F - r$; spread on collateral\\
$\gamma_K (t)$ & The cost of capital (the assets comprising the capital may themselves have a dividend yield and this can be incorporated into $\gamma_K (t)$)\\
$\gamma_E (t)$ & The effective tax rate\\
$\Delta\Vhat_B;\ \Delta\Vhat_C$ & Change in value of derivative on issuer default; on counterparty default\\
$\Delta_E ; \bar{\Delta}_E $ & Tax effect on counterparty default; Tax effect on counterparty default when $\psi=0$\\
$\epsilon_B$; $\epsilon_C$ & Hedging error on default of issuer. Sometimes split into terms independent of and dependent on capital $\epsilon_B = \epsilon_{B_0} + \epsilon_{B_K}$; hedging error on counterparty default\\
$P$ & $P = \alpha_1 P_1 + \alpha_2 P_2$ is the value of the own bond portfolio prior to default\\
$P_D$ & $P_D = \alpha_1 R_1 P_1 + \alpha_2 R_2 P_2$ is the value of the own bond portfolio after default\\
$\phi$ & Fraction of capital available for derivative funding\\
$\psi$ & Fraction of counterparty bond $P_C$ used in the hedge portfolio relative to full hedge\\
$\Gamma_C$ & Compensator of the counterparty jump\\
$m_{\lambda_C}$ & Market price of default risk for C\\\hline
\end{tabular}
\caption{\label{table:not}A summary of the notation, this is the same as Burgard and Kjaer (2013) and Green, Kenyon and Dennis (2014) apart from extensions.}
\end{table}

We extend the semi-replication arguments of \citet{Burgard2013a} and \citet*{Green2014b} to double semi-replication to take account of partial hedging of counterparty default, i.e. risk warehousing.  Thus we cover credit, funding, and capital.  For clarity we introduce Tax in the following section.  Notation is the same as  \citet*{Green2014b} with some additions (see Table \ref{table:not}). The sign convention is that the value of a cash amount is positive if received by the issuer. The aim is to find the economic or shareholder value of the derivative portfolio, $\hat{V}$. Note that here, as in \citet{Burgard2013a} and \citet*{Green2014b}, we neglect balance sheet feedback effects.

The derivation follows \citet*{Green2014b} with modifications for risk warehousing and the capital treatment of CVA hedges. The dynamics of the underlying assets are
\begin{align}
dS = & \mu_s S dt + \sigma_s S dW\\
dP_C = & r_C P_C dt - P_C dJ_C\\
dP_i = & r_i P_i dt - (1 - R_i)P_i dJ_B\quad\text{for}\quad i\in\{1,2\}
\end{align}
On default of the issuer, $B$, and the counterparty, $C$, the value of the derivative takes the following values
\begin{align}
\Vhat (t, S, 1, 0) = & g_B(M_B, X)\\
\Vhat (t, S, 0, 1) = & g_C(M_C, X),
\end{align}
where the two $g$ functions allow a degree of flexibility to be included in the model around the value of the derivative after default. The usual assumption is that
\begin{align}
g_B = &(V - X)^+ + R_B(V-X)^- + X\nonumber\\
g_C = &R_C(V-X)^+ + (V-X)^- + X,
\end{align}
where $x^+ =  \max\{x, 0\}$ and $x^- =  \min\{x, 0\}$.

We assume the funding condition (also known as the value equation in \cite{Duffie2001a}) for \Vhat:
\begin{equation}\label{eq:bondfunding}
\Vhat - X + \alpha_1 P_1 + \alpha_2 P_2 -\phi K = 0,
\end{equation}
where $\phi K$ represents the potential use of capital to offset funding requirements.
The growth in the cash account positions (gain equations in \cite{Duffie2001a}) associated with the stock and collateral (prior to rebalancing) are
\begin{align}
d\betaBar_S = & \delta (\gamma_S - q_S) S dt\\
d\betaBar_X = & -r_X X dt.
\end{align}
We assume, as in \cite{Burgard2013a} that the stock is funded by putting it in repo so that we only have dividend and borrow terms in $d\betaBar_S$.
Note that gains  come only from changes in the underlyings, e.g. interest on cash accounts, or dividends and borrow costs in repo, as the portfolio is self-financing (using the definition of self-financing from \cite{Duffie2001a}).

The  growth in the cash account, i.e. its gain equation, associated with the counterparty bond is now given by
\begin{equation}
\label{eq:betaC}d\bar{\beta}_C = -\psi\alpha_C q_C P_C dt - \Gamma_C dt.
\end{equation}
Note that we assume that the counterparty bond can be repo'd so we only have to pay the borrow cost $q_C$, as before.  Comparing this with the same expression in \citet{Burgard2013a} and \citet*{Green2014b} shows two changes:
\begin{itemize}
\item The bond holding is now $\psi\alpha_C$  where $\psi\in[0,1]$ reflects the fact that the hedge position will in general be partial.   $\alpha_C$ is the bond holding with a full hedge of the counterparty risk as previously.
\item A new term, $\Gamma_C$ has been introduced. This is the compensator for the losses that will now occur due to counterparty default when $\psi\ne1$.  This is the cashflow the bank requires to compensate for having a (partially) open risk position. However, we leave the form of $\Gamma_C$ unspecified for now. 
\end{itemize}
The degree of counterparty risk hedging is driven by $\psi$ where
\begin{align*}
\psi = 1 & \implies\quad\text{semi-replication}\\
\psi = 0 & \implies\quad\text{unhedged, i.e. risk is warehoused}
\end{align*}

The growth in the cash account (gain equation \cite{Duffie2001a}) associated with capital is also modified,
\begin{equation}
d\bar{\beta}_K = -\gamma_K(K^U - \psi K^R) dt.
\end{equation}
Here $K^U$ represents the capital requirement should no CVA hedging be performed. The capital relief associated with 100\% CVA hedging is given by $K^R$.  CVA VAR capital can be zero with allowable credit hedges.  However, when CDS are used, counterparty credit risk capital requirements are determined by the creditworthiness of the CDS provider (assuming this is better than the reference entity).  Delta hedging credit risk does not give a zero capital requirement for the same reason. We make the assumption that hedging through the counterparty bond $P_C$ is allowable for capital relief purposes in the CVA capital term under Basel III \cite{BCBS-189}. 
Further discussion on the general form of this term can be found in \citet{Green2014b}.

Applying It\^{o}'s lemma the change in the value of the derivative portfolio is
\begin{equation}
d\hat{V} = \frac{\partial \Vhat}{\partial t}dt + \frac{1}{2} \sigma^2 S^2 \frac{\partial^2\Vhat}{\partial S^2}  dt +\frac{\partial \Vhat}{\partial S} dS + \Delta \Vhat_B dJ_B + \Delta \Vhat_C dJ_C.
\end{equation}
Assuming the portfolio, $\Pi$, is self-financing i.e. its change in value comes only from gains in underlyings (stock, cash accounts, or bonds), gives
\begin{equation}
\begin{split}
d\Pi = & \delta dS + \delta (\gamma_S - q_S) S dt + \alpha_1 dP_1 + \alpha_2 dP_2 + \alpha^H_C dP_C \\
&- \alpha^H_C q_C P_C dt - r_X X dt - \gamma_K (K^U -\psi K^R)  dt - \Gamma_C dt.
\end{split}
\end{equation}
Adding the derivative and replicating portfolio together we obtain
\begin{align}
d\hat{V} + d\Pi = & \Bigg[\frac{\partial \Vhat}{\partial t} + \frac{1}{2} \sigma^2 S^2 \frac{\partial^2\Vhat}{\partial S^2} + \delta (\gamma_S - q_S) S + \alpha_1 r_1 P_1 + \alpha_2 r_2 P_2 \nonumber\\
&  + \alpha^H_C r_C P_C - \alpha^H_C q_C P_C - r_X X - \gamma_K (K^U - \psi K^R) - \Gamma_C \Bigg] dt\\\nonumber
& + \epsilon_B dJ_B + \epsilon_C dJ_C\\\nonumber
& + \left[\delta + \frac{\partial \Vhat}{\partial S}\right] dS,
\end{align}
where $\epsilon_B$ is the hedge error on the default of the issuer and $\epsilon_C$ is the hedge error on the default of the counterparty. 

Assuming replication of the derivative by the hedging portfolio, except on issuer or counterparty default gives,
\begin{equation}
d\hat{V} + d\Pi = 0. 
\end{equation}
The delta risk is eliminated as usual,
\begin{equation}
\delta = - \frac{\partial \hat{V}}{\partial S}.
\end{equation}
The hedging error on issuer default takes the same form as in \citet*{Green2014b} except that the capital relief has now been specified explicitly,
\begin{align}\label{eq:eBdef}
\epsilon_B = & \left[\Delta\hat{V}_B - (P - P_D) \right]\\\nonumber
= & g_B - X + P_D - \phi (K^U - \psi K^R)\\\nonumber
= & \epsilon_{B_0} + \epsilon_{B_K}
\end{align}
In the final line the hedging error has been split into a term which does not depend on capital, $\epsilon_{B_0}$ and a term which does depend on capital $\epsilon_{B_K}$. Note that $\epsilon_{B_K} \neq -\phi (K^U - \psi K^R)$ as the bond position is itself dependent on capital through the funding requirements as defined by equation \eqref{eq:bondfunding}. The hedging error on counterparty default is given by
\begin{align}
\epsilon_C = & \Delta\hat{V}_C - \psi \alpha_C P_C\\\nonumber
= & g_C - \hat{V} - \psi \alpha_C P_C\\\nonumber
= & (1 - \psi) (g_C - \hat{V}),
\end{align}
where we have used the fact that the original counterparty hedge was sized to remove all counterparty risk and hence $\alpha_C P_C = g_C - \hat{V}$.

We have not eliminated all sources of risk and hence we must choose how to deal with the residual risk to counterparty default. That is, what compensation does the bank require for running a partially open position?  There are many possibilities and we have chosen to price in the expected cost under the physical measure $\mathbb{P}$, i.e.
\begin{align}\label{eq:GammaC}
\Gamma_C = & - \mathbb{E}_t^{\mathbb{P}}[\epsilon_C dJ_C]\quad\forall\quad t \in [0,T]\\\nonumber
= & -(1 - \psi) (g_C - \Vhat) \lambda_C^{\mathbb{P}},
\end{align}
where $\lambda_C^{\mathbb{P}}$ is the hazard rate under the physical measure and hence is the instantaneous probability of a jump in this measure.

The hazard rate in the risk-neutral measure is $\lambda_C$ and, as with \citet{Burgard2013a}, we make the assumption that the counterparty bond can be repo'd at close to the risk-free rate and hence,
\begin{equation}
\lambda_C = r_C - q_C.
\end{equation} 
Furthermore we know that $\lambda_C$ must contain a market price of risk, $m_{\lambda_C}$ with respect to the physical probability of default and so,
\begin{align}
\lambda_C^{\mathbb{P}} = & \lambda_C  - m_{\lambda_C}\\\nonumber
= & \lambda_C(1 - \xi)
\end{align}
In general $m_{\lambda_C}$ may be rating (or level) dependent \cite{Hull2004b}.
Hence we have the following PDE for $\hat{V}$ under double semi-replication,
\begin{align}\label{eq:VhatPDE1}
0 = & \frac{\partial \Vhat}{\partial t} + \frac{1}{2} \sigma^2 S^2 \frac{\partial^2\Vhat}{\partial S^2} - (\gamma_S - q_S) S \frac{\partial \Vhat}{\partial S} - (r + \lambda_B + \tilde{\lambda}_C) \Vhat\nonumber\\
& + \tilde{\lambda}_C g_C  + \lambda_B g_B - \epsilon_B \lambda_B - s_X X - \gamma_K (K^U - \psi K^R) + r \phi (K^U - \psi K^R)\nonumber\\
& \hat{V}(T, S) = H(S).
\end{align}
where the bond funding equation \eqref{eq:bondfunding} has been used along with the yield of the issued bond, $r_i = r + (1- R_i) \lambda_B$ and the definition of $\epsilon_B$ in equation \eqref{eq:eBdef} to derive the result,
\begin{equation}
\alpha_1 r_1 P_1 + \alpha_2 r_2 P_2 = rX -(r + \lambda_B)\hat{V} -\lambda_B (\epsilon_B - g_B) + r \phi (K^U - \psi K^R),
\end{equation}
and we have defined $\tilde{\lambda}_C$ as
\begin{equation}\label{eq:tildelambda}
\tilde{\lambda}_C = \psi\lambda_C + (1-\psi)(1 - \xi)\lambda_C.
\end{equation}
$\tilde{\lambda}_C$ is the effective hazard rate in the presence of semi-replication of the counterpary risk. Equation \eqref{eq:tildelambda} shows that in the limits of full hedging and no hedging,
\begin{align*}
\psi = & 1\quad\tilde{\lambda}_C = \lambda_C\\
\psi = & 0\quad\tilde{\lambda}_C = \lambda_C^{\mathbb{P}}.
\end{align*}

The PDE \eqref{eq:VhatPDE1} has exactly the same structure as in \citet*{Green2014b} so writing $\hat{V} = V + U$ where $V$ satisfies the Black-Scholes equation allows a PDE to be written for $U$, the valuation adjustment,
\begin{align}
 \frac{\partial U}{\partial t} & + \frac{1}{2} \sigma^2 S^2 \frac{\partial^2 U}{\partial S^2} - (\gamma_S - q_S) S \frac{\partial U}{\partial S} - (r + \lambda_B + \tilde{\lambda}_C) U = \nonumber\\
& V \tilde{\lambda}_C - g_C \tilde{\lambda}_C + V \lambda_B - g_B \lambda_B + \epsilon_B \lambda_B + s_X X + (\gamma_K - r\phi)(K^U - \psi K^R)\nonumber\\
& U(T, S) = 0
\end{align}
Applying the Feynman-Kac theorem gives,
\begin{equation}
U = \text{CVA} + \text{DVA} + \text{FCA} + \text{COLVA} + \text{KVA},
\end{equation}
where
\begin{align}
\text{CVA} = & -\int_t^T \tilde{\lambda}_C(u) e^{-\int_t^u (r(s) + \lambda_B(s) + \tilde{\lambda}_C(s)) ds} \mathbb{E}_t \left[V(u) - g_C(V(u), X(u))\right] du\label{eq:intCVA}\\
\label{eq:intDVA}\text{DVA} = & -\int_t^T \lambda_B(u) e^{-\int_t^u (r(s) + \lambda_B(s) + \tilde{\lambda}_C(s)) ds}\mathbb{E}_t \left[V(u) - g_B(V(u), X(u))\right] du\\
\label{eq:intFCA}\text{FCA} = & -\int_t^T \lambda_B(u) e^{-\int_t^u (r(s) + \lambda_B(s) + \tilde{\lambda}_C(s)) ds}\mathbb{E}_t \left[\epsilon_{B_0}(u)  \right]du\\
\label{eq:intCOLVA}\text{COLVA} = & -\int_t^T s_X(u) e^{-\int_t^u (r(s) + \lambda_B(s) + \tilde{\lambda}_C(s)) ds} \mathbb{E}_t\left[X(u)\right] du\\
\label{eq:intKVA}\text{KVA} = & -\int_t^T e^{-\int_t^u (r(s) + \lambda_B(s) + \tilde{\lambda}_C(s)) ds} \nonumber\\
& \times \mathbb{E}_t \left[(\gamma_K (u) - r(u) \phi) (K^U - \psi K^R)(u)+ \lambda_B \epsilon_{B_K}(u)\right] du,
\end{align}
which, unsurprisingly, is identical with \citet*{Green2014b}, aside from the replacement of $\lambda_C$ with the effective hazard rate $\tilde{\lambda}_C$ and the explicit capital relief from hedging the CVA.  Note that complete credit hedging removes the CVA VAR capital requirement, but only changes the counterparty capital requirement to that for the provider of the credit hedge (if this is better rated).

If the market price of risk is positive so that the market implied probability of default is higher that that in the physical measure, $\tilde{\lambda}_C < \lambda_C$ and the CVA will be lower using this model, although the conditioning effect through $e^{-\int_t^u \tilde{\lambda}_C(s)) ds}$ in DVA, FCA, and COLVA will make these terms larger. The KVA term will increase because of the reduced CVA hedging benefit and because of the reduced conditioning effect. 

Since we have unhedged individual cashflows (although included via expectation) the portfolio $\Vhat+\Pi$ will show profits and losses.  We now turn to the potential tax consequences of these, and of the profits paying the return on capital.

\section{Tax Consequences and TVA}
The (possibly forced) choice to risk warehouse some fraction of the counterparty default risk means that there is a jump in value on counterparty default, and that there will be compensatory accruals prior to default. These effects lead to profits and losses which are taxable. Profits to pay shareholders' (or executives') required return on capital in KVA will also be subject to tax.  To cater for Taxation we extend the double semi-replication model above.  This leads to the introduction of a further valuation adjustment in XVA, \emph{TVA, Taxation Valuation Adjustment}.

We apply a single effective tax rate to profits or losses as they occur so that tax losses lead to tax credits.   We assume that the bank has sufficient profits that any tax credits can be used. The possibility of not using tax credits over an extended period is considered to be part of the bank default probability.

To add taxation to the replication portfolio we introduce a further cash account associated with tax. It has two terms, reflecting the tax rate paid on the tax liability and the tax effect on counterparty default (prior to rebalancing).
\begin{equation}
d\betaBar_E = -\gamma_E (t) E(t) dt + \Delta_E dJ_C,
\end{equation}
Again here there is no term in $dJ_B$ as we assume tax effects on own default are incorporated into the recovery rate $R_B$.  Note that this account has two underlyings: a tax liability (or credit) $E(t)$ on non-jump times; and a tax liability (or credit) effect at counterparty default.

The change in the replicating portfolio is now given by,
\begin{equation}\label{eq:dPiE}
\begin{split}
d\Pi = & \delta dS + \delta (\gamma_S - q_S) S dt + \alpha_1 dP_1 + \alpha_2 dP_2 + \psi\alpha_C dP_C - \psi\alpha_C q_C P_C dt\\
& - r_X X dt - \gamma_K (K^U -\psi K^R)  dt - \Gamma_C dt -\gamma_E (t) E dt + \Delta_E dJ_C.
\end{split}
\end{equation}
Combining the change in the derivative, which is unchanged, with equation \eqref{eq:dPiE} gives,
\begin{align}
d\Vhat + d\Pi = & \Bigg[\frac{\partial \Vhat}{\partial t} + \frac{1}{2} \sigma^2 S^2 \frac{\partial^2\Vhat}{\partial S^2} + \delta (\gamma_S - q_S) S + \alpha_1 r_1 P_1 + \alpha_2 r_2 P_2 \nonumber\\
&  + \psi\alpha_C r_C P_C - \psi\alpha_C q_C P_C - r_X X - \gamma_K (K^U - \psi K^R) - \Gamma_C -\gamma_E (t) E\Bigg] dt\\\nonumber
& + \epsilon_B dJ_B + \epsilon_C dJ_C\\\nonumber
& + \left[\delta + \frac{\partial \Vhat}{\partial S}\right] dS,
\end{align}
where the hedging error on own default is unchanged but the hedging error on counterparty default is given by
\begin{align}
\epsilon_C = & \Delta\Vhat_C - \psi\alpha_C P_C + \Delta_E\\\nonumber
= & (1 - \psi) (g_C - \Vhat + \bar{\Delta}_E),
\end{align}
where we have assumed that the tax effect on default of the counterparty scales linearly with the proportion of hedging $\Delta_E = (1 - \psi)\bar{\Delta}_E$. 

As in equation \eqref{eq:GammaC} we choose the compensator to match the expected cost of counterparty default in the physical measure,
\begin{equation}
\Gamma_C = -(1 - \psi) (g_C - \Vhat + \bar{\Delta}_E) \lambda_C(1 - \xi).
\end{equation}
Hence the PDE for $\hat{V}$ becomes,
\begin{align}\label{eq:VhatPDE}
0 = & \frac{\partial \Vhat}{\partial t} + \frac{1}{2} \sigma^2 S^2 \frac{\partial^2\Vhat}{\partial S^2} - (\gamma_S - q_S) S \frac{\partial \Vhat}{\partial S} - (r + \lambda_B + \tilde{\lambda}_C) \Vhat\nonumber\\
& + \tilde{\lambda}_C g_C  + \lambda_B g_B - \epsilon_B \lambda_B - s_X X - (\gamma_K - r\phi) (K^U - \psi K^R)\nonumber\\
& -\gamma_E (t) E - \lambda_C(1 - \xi) (1 - \psi) \bar{\Delta}_E\nonumber\\
& \Vhat(T, S) = H(S).
\end{align}
Finally the PDE for $U$ becomes,
\begin{align}
\frac{\partial U}{\partial t} & + \frac{1}{2} \sigma^2 S^2 \frac{\partial^2 U}{\partial S^2} - (\gamma_S - q_S) S \frac{\partial U}{\partial S} - (r + \lambda_B + \tilde{\lambda}_C) U = \nonumber\\
& V \tilde{\lambda}_C - g_C \tilde{\lambda}_C + V \lambda_B - g_B \lambda_B + \epsilon_B \lambda_B + s_X X + (\gamma_K - r\phi)(K^U - \psi K^R)\nonumber\\
& -\gamma_E (t) E - \lambda_C(1 - \xi) (1 - \psi) \bar{\Delta}_E\nonumber\\
& U(T, S) = 0
\end{align}
Again applying the Feynman-Kac theorem yields,
\begin{equation}
U = \text{CVA} + \text{DVA} + \text{FCA} + \text{COLVA} + \text{KVA} + \text{TVA},
\end{equation}
where
\begin{align}
\text{CVA} = & -\int_t^T \tilde{\lambda}_C(u) e^{-\int_t^u (r(s) + \lambda_B(s) + \tilde{\lambda}_C(s)) ds} \mathbb{E}_t \left[V(u) - g_C(V(u), X(u))\right] du\label{e:cvaGeneral}\\
\text{DVA} = & -\int_t^T \lambda_B(u) e^{-\int_t^u (r(s) + \lambda_B(s) + \tilde{\lambda}_C(s)) ds}\mathbb{E}_t \left[V(u) - g_B(V(u), X(u))\right] du
\label{e:dvaGeneral}\\
\text{FCA} = & -\int_t^T \lambda_B(u) e^{-\int_t^u (r(s) + \lambda_B(s) + \tilde{\lambda}_C(s)) ds}\mathbb{E}_t \left[\epsilon_{B_0}(u)  \right]du
\label{e:fcaGeneral}\\
\text{COLVA} = & -\int_t^T s_X(u) e^{-\int_t^u (r(s) + \lambda_B(s) + \tilde{\lambda}_C(s)) ds} \mathbb{E}_t\left[X(u)\right] du
\label{e:colvaGeneral}\\
\text{KVA} = & -\int_t^T e^{-\int_t^u (r(s) + \lambda_B(s) + \tilde{\lambda}_C(s)) ds} \nonumber\\
& \times \mathbb{E}_t \left[(\gamma_K (u) - r(u) \phi) (K^U - \psi K^R)(u)+ \lambda_B \epsilon_{B_K}(u)\right] du
\label{e:kvaGeneral}\\
\text{TVA} = & -\int_t^T e^{-\int_t^u (r(s) + \lambda_B(s) + \tilde{\lambda}_C(s)) ds} \nonumber\\
& \times \mathbb{E}_t \left[\gamma_E (t) E + \lambda_C(1 - \xi) (1 - \psi) \bar{\Delta}_E\right] du
\label{e:tvaGeneral}
\end{align}
Capital terms are specified by capital Regulations, similarly tax terms are specified by tax Laws.  Tax terms at non-jump times and on counterparty default are the direct result of profits and losses, so\footnote{The authors are not tax experts so this is for illustration only.}: 
\bean
E &=& \gamma_K (u)  (K^U - \psi K^R)(u)+ \I{B} \lambda_B( \epsilon_{B_K}(u)  +  \epsilon_{B_0}(u)),  \label{e:E}\\
\DeltaBar_E &=& {}- \gamma_E (u)\left(V(u) - g_C(V(u), X(u)) \right)  \label{e:deltaE1}
\eean
The TVA term itself states that more profits must be made so that, after paying tax, the required return on capital to shareholders can be met.  It also states that more profits must be made so that the required return on open risk can be made by the bank (according to the choice of compensator $\Gamma$). An alternative way of looking at the TVA term is that it states that the value to shareholders of $\Vhat$ is reduced because taxes must be paid.  
\begin{itemize}
	\item Equation \ref{e:E} states that profits to pay for capital are taxed.  Note that the source of the profit (whether it comes from returns on existing capital or not) is irrelevant, hence the $-r(u)\phi$ term is not present.
	\item The indicators $\I{B}$ in Equation \ref{e:E} describe the choice of the tax authority on cashflows accrued relative to own-default.  If the tax authority follows the accounting treatment of, say, FAS157 or IFRS13 that the correct price (from the point of view of the tax authority) includes own-creditworthiness then $\I{B}=1$, otherwise $\I{B}=0$.  In other words, does the tax authority tax the accruals that cancel out PnL bleed?  That is, does the tax authority look on the bank as a going concern (so chooses $\I{B}=0$)?
	\item Equation \ref{e:deltaE1} states that the bank requires compensation for the expected tax effects of the value jump on counterparty default.  This term could, alternatively, be included in the CVA term.
\end{itemize}
  
Hence a new valuation adjustment, TVA, has now been added to the existing set of adjustments present earlier under semi-replication.

\section{Numerical Examples}

Here we provide examples based on 10Y GBP interest rate swaps (IRS) to demonstrate the effect of risk warehousing on valuation adjustments, and to show the magnitude of tax valuation adjustments.  We follow 'strategy 1' in \cite{Burgard2013a} so that there is no shortfall on own default but any potential windfall after default is not monetized, i.e. the spread between the two own bonds is not exploited.  Equations \ref{e:cvaGeneral}--\ref{e:tvaGeneral} specialize to:
\begin{align}
\text{CVA} = & -(1-R_C)\int_t^T \tilde{\lambda}_C(u) e^{-\int_t^u (\lambda_B(s) + \tilde{\lambda}_C(s)) ds} \mathbb{E}_t \left[ e^{-\int_t^u (r(s)ds} V(u)^+ \right] du\label{e:cvaS1}\\
\text{DVA} = & -(1-R_B)\int_t^T \lambda_B(u) e^{-\int_t^u ( \lambda_B(s) + \tilde{\lambda}_C(s)) ds}\mathbb{E}_t \left[e^{-\int_t^u (r(s)ds} V(u)^- \right] du
\label{e:dvaS1}\\
\text{FCA} = & -(1-R_B)\int_t^T \lambda_B(u) e^{-\int_t^u ( \lambda_B(s) + \tilde{\lambda}_C(s)) ds}\mathbb{E}_t \left[e^{-\int_t^u (r(s)ds} V(u)^+  \right]du
\label{e:fcaS1}\\
\text{KVA} = & -\int_t^T e^{-\int_t^u ( \lambda_B(s) + \tilde{\lambda}_C(s)) ds} \nonumber\\
& \times \mathbb{E}_t \left[e^{-\int_t^u (r(s)ds} (\gamma_K (u) - r(u) \phi) (K^U - \psi K^R)(u)\right] du
\label{e:kvaS1}\\
\text{TVA} = & -\int_t^T e^{-\int_t^u ( \lambda_B(s) + \tilde{\lambda}_C(s)) ds}  \mathbb{E}_t \left[e^{-\int_t^u (r(s)ds} \right. \nonumber\\
&\left. \vphantom{ e^{-\int_t^u}} 
\gamma_E (u)\left( \gamma_K (u)  (K^U - \psi K^R)(u) - \lambda_C(1-R_C)(1 - \xi) (1 - \psi) V(u)^+\right)\right] du
\label{e:tvaS1} 
\end{align}
Since we use interest rate swaps we have stochastic interest rates and so the $r(u)$ discounting is within the expectations.  The derivation follows the same steps as for derivatives of stock prices (see \cite{Green2014b} for the KVA case).

The 10Y GBP swaps have semi-annual payment schedules and fixed rate of 2.7\%.  Issuer credit spread is flat 100bps with 40\%\ recovery.  We assume there is no capital priced in to the credit spread.  Counterparty data is shown in Table \ref{tab:ctpyspread}.

Capital requirements are as in \cite{Green2014b}: we use the standardised approach for Market Risk capital ; current exposure method for Exposure at Default in Counterparty Credit Risk (CCR) capital with standardised approach for weights;  CVA VAR capital uses standardized approach with the approximation for large numbers of counterparties.  Cost of capital, $\gamma_K$, is set at 10\%\ and minimum capital 8\%.

Tax rate, $\gamma_E$, is set to 21\%\ and applied to cash flows as they occur, including counterparty default.

\begin{table}[ht]
\centering
\begin{tabular}{p{3cm}|r|p{3cm}|p{3cm}}
{\bf Counterparty Rating} & {\bf CDS (bp)} & {\bf Standardized Risk Weight} &{\bf CVA Risk Weight $w_i$}\\\hline
AAA & 30 & 20\% & 0.7\%\\
A & 75 & 50\% & 0.8\%\\
BB & 250 & 100\% & 2\%\\
CCC & 750 & 150\% & 10\%\\\hline
\end{tabular}
\caption{\label{tab:ctpyspread}Counterparty data for the examples.}
\end{table}

\paragraph{Base Case: back-to-back IRS with CCDS on uncollateralized side} Base case is a pair of back-to-back 10Y GBP IRS, one fully collateralized and the other uncollateralized.  We assume that a (fully collateralized) CCDS is available to perfectly credit hedge the uncollateralized IRS and that the CVA charge is used to purchase it.  The CCDS is a contingent credit default swap that perfectly matches the exposure of the uncollateralized IRS under all states of the world up to maturity of the IRS.  The exposure of the uncollateralized IRS is the value multiplied by the loss given default.  Thus we have a perfect credit protection instrument, but this does not reduce CCR capital to zero it just switches to calculation to the rating of the CDS seller.

\begin{table}[htbp]
	\centering
\resizebox{1.2\columnwidth}{!}{%
			\begin{tabular}{p{1cm}p{1cm}p{1cm}|c|c|ccc|ccc|c|c}
  \multicolumn{3}{c|}{\bf Credit Hedge} & $\phi$ & {\bf c/p} & \text{\bf CVA} & \text{\bf DVA}& \text{\bf FCA}&\multicolumn{3}{|c|}{\bf KVA} & \text{\bf TVA} & \text{\bf Total}\\\hline
 \text{Source} & $\psi$ & $m_{\lambda_C}$ &  & \text{Rating} & &  &  & \text{MR} & \text{CCR} & \text{CVA} &  &  \\
 \text{A} & 1 & \text{na} & 0 & \text{AAA} & -4 & 39 & -14 & 0 & -3 & 0 & -1 & 17 \\
 \text{A} & 1 & \text{na} & 0 & \text{A} & -10 & 38 & -14 & 0 & -8 & 0 & -2 & 5 \\
 \text{A} & 1 & \text{na} & 0 & \text{BB} & -31 & 33 & -12 & 0 & -7 & 0 & -1 & -18 \\
 \text{A} & 1 & \text{na} & 0 & \text{CCC} & -68 & 24 & -9 & 0 & -5 & 0 & -1 & -60 \\ \hline
 \text{A} & 1 & \text{na} & 1 & \text{AAA} & -4 & 39 & -14 & 0 & -2 & 0 & 0 & 19 \\
 \text{A} & 1 & \text{na} & 1 & \text{A} & -10 & 38 & -14 & 0 & -4 & 0 & -1 & 9 \\
 \text{A} & 1 & \text{na} & 1 & \text{BB} & -31 & 33 & -12 & 0 & -4 & 0 & -1 & -14 \\
 \text{A} & 1 & \text{na} & 1 & \text{CCC} & -68 & 24 & -9 & 0 & -3 & 0 & -1 & -57 \\ \hline
\end{tabular}%
}
	\caption{Base case: back to back 10Y GBP interest rate PAY swaps with different counterparties, one side collateralized the other uncollateralized; with exact credit hedge ($\psi=1$) on uncollateralized side.  $\phi$ states whether income on capital is included in the cost of capital.  Values for XVA are bps of notional.}
	\label{t:base}
\end{table}

The valuation adjustments are shown in Table \ref{t:base} for a range of counterparty ratings for the uncollateralized side.    In particular:
\begin{itemize}
	\item zero Market Risk capital cost because the IRS are back-to-back
	\item zero CVA VAR capital cost because we are collateralized on one side and have credit protection from the CCDS on the other
	\item tax cost is close to zero because the capital costs are low and there is no value jump on counterparty default
	\item sign of the total XVA adjustment changes from positive to negative with increasing riskiness of counterparty
\end{itemize}

\paragraph{Cases with warehoused credit risk} are shown in Table \ref{t:neg} and Table \ref{t:pos} when there is no credit hedge on the uncollateralized side of the back-to-back IRS swaps.  Table \ref{t:pos} shows results with a positive market price of risk, e.g. during or shortly after a period of perceived high credit risk, so the market-implied hazard rate is higher than the physical-measure hazard rate.  Table \ref{t:neg} shows the opposite situation, e.g. after a long period of perceived low credit risk.

\paragraph{Effect of Risk Warehousing} We can observe the following points:
\begin{itemize}
	\item CVA, DVA, and FCA change by roughly 50\%--25\%\ relative to the base case.  The direction depends on the sign of the market price of default risk as it expresses the difference between the risk-neutral and physical measure hazard rates
	\item CCR capital costs roughly double because there is no credit protection
	\item CVA VAR capital costs appear and vary by up to roughly 25\%\ depending on the sign of the market price of default risk
\end{itemize}

\paragraph{Effects of Tax}
 Tax effects are highly dependent on the sign of the market price of credit risk.  This is because there is an opposition between tax effects on counterparty default and tax effects of capital.  This opposition is almost balanced here when the market is underestimating the physical probability of default, i.e. negative market price of jump to default risk.  When the market overestimates the hazard rate then there is relatively little tax gain from any losses on counterparty default.  The opposite is true when the market underestimates the hazard rate.

All effects are reduced when income from capital is used to reduce the cost of capital.  Note that tax is taken from the whole cost of capital as the source of the profits is irrelevant to the tax authority.

\paragraph{Market price of jump to default risk}  We have considered positive and negative market prices of jump to default risk, $m_{\lambda_C}$.  However, if capital relief is priced into observed CDS spreads this will mimic the effect of a positive $m_{\lambda_C}$.  This suggests a future systematic bias towards (effective) positive prices $m_{\lambda_C}$.   In our examples it would be very roughly 0.5.

\begin{table}[htbp]
	\centering
\resizebox{1.2\columnwidth}{!}{%
			\begin{tabular}{p{1cm}p{1cm}p{1cm}|c|c|ccc|ccc|c|c}
  \multicolumn{3}{c|}{\bf Credit Hedge} & $\phi$ & {\bf c/p} & \text{\bf CVA} & \text{\bf DVA}& \text{\bf FCA}&\multicolumn{3}{|c|}{\bf KVA} & \text{\bf TVA} & \text{\bf Total}\\\hline
 \text{Source} & $\psi$ & $m_{\lambda_C}$ &  & \text{Rating} & &  &  & \text{MR} & \text{CCR} & \text{CVA} &  &  \\
 \text{A} & 0 & 0.5 & 0 & \text{AAA} & -2 & 40 & -14 & 0 & -3 & -9 & -2 & 10 \\
 \text{A} & 0 & 0.5 & 0 & \text{A} & -5 & 39 & -14 & 0 & -8 & -10 & -3 & 0 \\
 \text{A} & 0 & 0.5 & 0 & \text{BB} & -17 & 37 & -13 & 0 & -15 & -23 & -4 & -36 \\
 \text{A} & 0 & 0.5 & 0 & \text{CCC} & -43 & 31 & -11 & 0 & -19 & -103 & -17 & -162 \\ \hline
 \text{A} & 0 & 0.5 & 1 & \text{AAA} & -2 & 40 & -14 & 0 & -2 & -6 & -1 & 14 \\
 \text{A} & 0 & 0.5 & 1 & \text{A} & -5 & 39 & -14 & 0 & -4 & -7 & -1 & 8 \\
 \text{A} & 0 & 0.5 & 1 & \text{BB} & -17 & 37 & -13 & 0 & -8 & -16 & -2 & -19 \\
 \text{A} & 0 & 0.5 & 1 & \text{CCC} & -43 & 31 & -11 & 0 & -10 & -74 & -9 & -116 \\ \hline
\end{tabular}%
}
	\caption{No credit hedge, and positive market price of risk, e.g. during or shortly after a period of perceived high credit risk.  $\phi$ states whether income on capital is included in the cost of capital.  Values for XVA are bps of notional.}
	\label{t:pos}
\end{table}

\begin{table}[htbp]
	\centering
\resizebox{1.2\columnwidth}{!}{%
			\begin{tabular}{p{1cm}p{1cm}p{1cm}|c|c|ccc|ccc|c|c}
  \multicolumn{3}{c|}{\bf Credit Hedge} & $\phi$ & {\bf c/p} & \text{\bf CVA} & \text{\bf DVA}& \text{\bf FCA}&\multicolumn{3}{|c|}{\bf KVA} & \text{\bf TVA} & \text{\bf Total}\\\hline
 \text{Source} & $\psi$ & $m_{\lambda_C}$ &  & \text{Rating} & &  &  & \text{MR} & \text{CCR} & \text{CVA} &  &  \\
 \text{A} & 0 & -0.5 & 0 & \text{AAA} & -6 & 39 & -14 & 0 & -3 & -8 & -1 & 6 \\
 \text{A} & 0 & -0.5 & 0 & \text{A} & -15 & 37 & -13 & 0 & -7 & -9 & 0 & -9 \\
 \text{A} & 0 & -0.5 & 0 & \text{BB} & -43 & 31 & -11 & 0 & -13 & -21 & 2 & -55 \\
 \text{A} & 0 & -0.5 & 0 & \text{CCC} & -84 & 19 & -7 & 0 & -13 & -75 & -1 & -161 \\  \hline
 \text{A} & 0 & -0.5 & 1 & \text{AAA} & -6 & 39 & -14 & 0 & -2 & -6 & 0 & 11 \\
 \text{A} & 0 & -0.5 & 1 & \text{A} & -15 & 37 & -13 & 0 & -4 & -7 & 1 & -1 \\
 \text{A} & 0 & -0.5 & 1 & \text{BB} & -43 & 31 & -11 & 0 & -7 & -15 & 4 & -40 \\
 \text{A} & 0 & -0.5 & 1 & \text{CCC} & -84 & 19 & -7 & 0 & -7 & -55 & 5 & -130 \\ \hline
\end{tabular}%
}
	\caption{No credit hedge, and negative market price of risk, e.g. after a long period of low perceived credit risk.  $\phi$ states whether income on capital is included in the cost of capital.  Values for XVA are bps of notional.}
	\label{t:neg}
\end{table}

\section{Conclusions}

Limited CDS markets naturally restrict the ability to obtain credit protection and force institutions to warehouse credit risk.  This will produce profits and losses which may be taxable.  In addition the profits to pay the return on capital to shareholders may be taxable.  Thus we have introduced double-semi-replication to tackle limited credit protection and a Tax Valuation Adjustment to formalize tax effects.  We find highly significant effects of warehousing credit risk on existing XVA elements and on TVA.  These effects are dominated by the sign of the market price of jump to default risk.  This sign may have a systematic positive bias, i.e. towards increased market implied hazard rates in as much as capital relief is priced in to CDS spreads.  A positive market price of default risk means that CVA, DVA, and FCA are generally reduced but the capital and tax elements are increased because there is less actual probability of default.

\section*{Acknowledgements}

The authors gratefully acknowledge useful discussions with Lincoln Hannah and Prof. Damiano Brigo.

\bibliographystyle{plainnat}
\bibliography{kenyon_general}

\end{document}